\documentclass[3p, preprint,12pt,fleqn]{elsarticle}

\usepackage{hyperref}
\usepackage[latin9]{inputenc}
\usepackage{amsmath}
\usepackage{tabularx}
\usepackage{multirow}
\usepackage{rotating}
\usepackage{color}
\usepackage{babel}
\usepackage{array}
\usepackage{float}
\usepackage{slashed}
\usepackage{booktabs}

\journal{Journal of \LaTeX\ Templates}

\def\bt{\begin{equation}}
\def\bea{\begin{eqnarray}}
\def\ee{\end{equation}}
\def\eea{\end{eqnarray}}

\begin{document}

\begin{frontmatter}

\title{Probing the Tau Neutrino Magnetic Moment via the $\mu^{-}\gamma \rightarrow \mu^{-}\nu\bar{\nu}$ Process at Future Muon Colliders} 


\author[myfirstaddress]{D. Y{\i}lmaz}

\address[myfirstaddress]{Physics Engineering Department, Ankara University, 06100, Ankara, T\"{u}rkiye }
\ead{dyilmaz@eng.ankara.edu.tr}

\author[mymainaddress,mysecondaryaddress]{M. \c{S}ahin\corref{mycorrespondingauthor}}
\cortext[mycorrespondingauthor]{Corresponding author}
\address[mymainaddress]{Engineering and Natural Siciences Faculty, Department of Computer Engineering, Usak University, 64200, U\c{s}ak, T\"{u}rkiye}
\address[mysecondaryaddress]{Energy, Environment and Sustainability Application and \\Research Center, 64200, U\c{s}ak University, U\c{s}ak, T\"{u}rkiye}
\ead{mehmet.sahin@usak.edu.tr}

\author[myaddress]{\.{I}. \c{S}ahin}

\address[myaddress]{Physics Department, Ankara University, 06100, Ankara, T\"{u}rkiye }
\ead{inancsahin@ankara.edu.tr}



\begin{abstract}
We investigate the sensitivity of future muon colliders to the anomalous magnetic moment of the tau neutrino within a model-independent effective field theory framework. The analysis is performed through the subprocess $\mu^{-}\gamma \rightarrow \mu^{-}\nu\bar{\nu}$ where the initial-state photon is described using the Equivalent Photon Approximation. Signal and background events are generated with MadGraph5$\_$aMC$@$NLO, followed by parton showering and hadronization with PYTHIA8. Detector effects are simulated using DELPHES3, while the event reconstruction and analysis are carried out via the MadAnalysis5 framework. Optimized kinematic selection criteria are determined for center-of-mass energies of 3, 10 and 30 TeV muon colliders. The expected sensitivities are evaluated using the Asimov significance method. The best sensitivity is obtained for the 30 TeV muon collider option, where the anomalous magnetic moment of the tau neutrino can be probed down to  $\mu_{\tau\tau}=5.21\times10^{-8}\,\mu_B$ at the 5$\sigma$ significance level assuming a 1\% systematic uncertainty. Our sensitivity bounds improve the current direct experimental limit reported by the DONUT Collaboration up to one order of magnitude and demonstrate the strong potential of future muon colliders for probing neutrino electromagnetic interactions.
\end{abstract}


\end{frontmatter}


\section{Introduction}

The discovery of neutrino oscillations \cite{Cleveland:1998nv, Super-Kamiokande:2001ljr, SNO:2001kpb,KamLAND:2002uet} has been one of the strongest experimental pieces of evidence demonstrating the incompleteness of the Standard Model (SM) and marked the beginning of a new era in particle physics. The oscillation phenomenon revealed that neutrinos possess non-zero masses, demonstrating that the massless neutrino approximation predicted by the SM is invalid. Therefore, the origin of neutrino masses, the structure of lepton flavor mixing, and possible non-standard interactions of neutrinos are among the most important research topics in modern particle physics. In particular, the electromagnetic properties of neutrinos, while highly suppressed within the SM, are considered one of the most sensitive indicators in the search for new physics, as they can be significantly enhanced in many new physics scenarios.

In the minimal extension of the SM including Dirac neutrinos, the neutrino magnetic moment is generated at the one-loop level and scales linearly with the neutrino mass. In this case, the neutrino magnetic moment is given by \cite{Marciano:1977wx, Lee:1977tib, Fujikawa:1980yx}:
\begin{equation}
\mu_{\nu}=
\frac{3eG_Fm_{\nu}}
{8\sqrt{2}\pi^2}
\simeq
3.2\times10^{-19}
\left(
\frac{m_{\nu}}
{1~{\rm eV}}
\right)
\mu_B
\end{equation}

This prediction lies about seven to eight orders of magnitude below current experimental sensitivities. Therefore, observing a neutrino magnetic moment significantly above the SM expectation would be a strong indicator of the existence of new particles or new interaction mechanisms. Indeed, many approaches beyond the SM, including left-right symmetric models, supersymmetry, extra-dimensional scenarios, and additional scalar or vector bosons, can increase the neutrino magnetic moment to experimentally attainable levels \cite{Czakon:1998rf, Gozdz:2006iz, McLaughlin:2000zf, Sahdev:2024wie}. Therefore, tighter experimental limits on the neutrino magnetic moment play a crucial role in testing new physics models.

The electromagnetic properties of neutrinos are relevant to a broad range of studies in particle physics, astrophysics, and cosmology. A non-zero magnetic moment can alter the energy loss mechanisms of stars, affect the evolution of white dwarf and red giant stars, play a role in the dynamics of supernova explosions, and modify neutrino emission under strong magnetic fields \cite{Giunti:2014ixa}. Therefore, limits obtained from laboratory experiments are evaluated together with results obtained from astrophysical observations, providing complementary information.
 
Neutrino-electron elastic scattering experiments using reactors and solar neutrinos, along with astrophysical observations, provide the strongest experimental limits currently available on the neutrino magnetic moment. Current laboratory and astrophysical data indicate that the neutrino magnetic moment cannot be greater than approximately $10^{-11}\,\mu_B$--$10^{-12}\,\mu_B$ \cite{Raffelt:1990pj, Miranda:2003yh, Allen:1992qe, LSND:2001akn, DONUT:2001zvi, PandaX-II:2020udv, Borexino:2008dzn, Capozzi:2020cbu, Vassh:2015yza, Grohs:2023xwa}. It should be noted that these limits correspond to the effective neutrino magnetic moment ($\mu_{\rm eff}$) measured in neutrino-electron scattering experiments and reflect the combined effect of neutrino mixing and elements of the magnetic moment matrix. In contrast, direct experimental constraints on the tau neutrino magnetic moment remain several orders of magnitude weaker due to the experimental difficulties in generating and detecting $\nu_\tau$ beams. Therefore, while translating limits on effective magnetic moments into constraints on individual matrix elements is generally model-dependent, the diagonal magnetic moment of the tau neutrino is significantly less constrained than the corresponding electron and muon neutrino diagonal moments by direct measurements. Currently, the strongest direct laboratory limit for the tau neutrino is reported by the DONUT Collaboration \cite{DONUT:2001zvi} as
\[
\mu_{\nu_\tau}<3.9\times10^{-7}\,\mu_B
\]
(90 \%C.L.). This significant difference in sensitivity makes the investigation of the magnetic moment of the tau neutrino in future high-energy colliders important for the pursuit of new physics. 

Various phenomenological studies \cite{Sahin:2010zr, Sahin:2012zm, Senol:2012sn} have examined the potential of proposed electron-positron, and hadron colliders to improve the sensitivity to the magnetic moment of the tau neutrino. Analyses using the effective field theory approach, in particular, are widely accepted because they allow for the model-independent investigation of anomalous electromagnetic couplings. Previous analyses, particularly those conducted in hadron and lepton colliders, have shown that the sensitivities to the magnetic moment of the tau neutrino remain below the current direct experimental limits. However, these studies demonstrate that precise measurements performed under high energy and high luminosity conditions can significantly improve current experimental limits. Therefore, future muon colliders operating at the multi-TeV scale have the potential to investigate the electromagnetic properties of the tau neutrino with substantially improved sensitivity.

More recently, the electromagnetic properties of neutrinos at future muon colliders have been analyzed through a model-independent effective field theory framework \cite{DYilmaz:2026}. In that study, anomalous $\nu\bar{\nu}\gamma\gamma$ interactions parametrized by dimension-7 effective operators were analyzed through the $\mu^{-}\gamma\rightarrow\mu^{-}\nu\bar{\nu}$ subprocess at center-of-mass energies of 3, 10, and 30 TeV. The results demonstrated the strong sensitivity of high-energy muon colliders to anomalous neutrino electromagnetic interactions. Motivated by these results, the present work investigates the neutrino magnetic moment in the same high-energy muon-collider environment using the same subprocess, with particular emphasis on the anomalous magnetic moment of the tau neutrino.

Muon colliders, whose preliminary studies began in the late 1960s \cite{Tikhonin:2008pw}, have emerged as a promising facility for future high-energy physics studies. Owing to the much larger mass of muons compared with electrons, synchrotron radiation is strongly suppressed at a given beam energy. This capability allows muon colliders to achieve multi-TeV energy levels. In recent years, different collider energy scenarios extending to 3, 10, and 30 TeV have been proposed by initiatives such as the Muon Accelerator Program (MAP) and the International Muon Collider Collaboration, making high-energy muon colliders an attractive facility for exploring physics beyond the SM \cite{Boscolo:2018ytm, Delahaye:2019omf, InternationalMuonCollider:2024jyv}.

However, the short lifetime of muons presents significant technical challenges in terms of obtaining high-luminosity muon beams and maintaining beam quality. In this context, the experimental demonstration of ionization cooling by the Muon Ionization Cooling Experiment (MICE) collaboration has been a significant development for the realization of future muon colliders \cite{MICE:2019jkl}.

One more important issue for muon colliders is the beam-induced background (BIB) that is generated by the decay processes of muons. Secondary particles resulting from the decay of muons circulating in the accelerator can create additional backgrounds in the detector environment. Therefore, BIB effects must be considered in both detector design and physics analysis. On the other hand, thorough investigations into simulation and detector design demonstrate that BIB effects can be substantially lessened by implementing appropriate shielding, timing techniques, and well-optimized geometrical configurations of detectors \cite{MuonCollider:2022ded, MuonCollider:2022glg, Bartosik:2020xwr, Collamati:2021sbv, Lu:2020dkx}. Thus, even with the technical difficulties faced, muon colliders continue to be promising options for research extending beyond the SM in physics.

This study investigates the sensitivity of future muon colliders to the anomalous magnetic moment of the tau neutrino using a model-independent effective field theory description. The analysis is performed through the subprocess $\mu^{-}\gamma \rightarrow \mu^{-}\nu\bar{\nu}$ for $\sqrt{s}$=3, 10, and 30 TeV muon collider energy options.

The photon participating in the subprocess is considered as a  Weizs\"acker--Williams equivalent photon \cite{vonWeizsacker:1934nji, Williams:1934ad} emitted from a high-energy muon beam. The Equivalent Photon Approximation (EPA) is based on treating low-virtuality photons emitted by highly energetic charged particles as an effective beam of real photons and is widely employed in the theoretical and phenomenological analysis of photon-induced processes at high-energy colliders. 

Signal and background processes were analyzed using detailed Monte Carlo simulations, optimized event selection criteria were applied, and expected sensitivities were determined using Asimov significance method. The obtained sensitivities are compared with the current experimental bounds and previous phenomenological studies in order to assess the potential of future muon colliders for probing the magnetic moment of the tau neutrino.

The remainder of this paper is organized as follows. Section II presents the theoretical framework. Section III describes the signal and background analysis. Section IV contains the statistical analysis and numerical results. Finally, Section V summarizes our conclusions.

\section{Theoretical Framework}
\subsection{Effective $\nu\bar{\nu}\gamma$ Interaction}
In the minimal extension of the SM including Dirac neutrinos, the electromagnetic interactions of Dirac neutrinos can be described by an effective dipole operator. The most general dimension-5 effective Lagrangian describing neutrino-photon interactions, consistent with Lorentz invariance, $SU(2)_L \times U(1)_Y$ gauge invariance, and lepton number conservation, can be written as \cite{Larios:1995np, Maya:1998ee, Bell:2005kz, Larios:2002gq}
\begin{equation}
\mathcal{L}_{\nu\bar{\nu}\gamma}
=
\frac{1}{2}
\mu_{ij}
\,
\bar{\nu}_i
\sigma^{\mu\nu}
\nu_j
F_{\mu\nu},
\label{eq:Lmag}
\end{equation}
where $F_{\mu\nu}$ is the electromagnetic field tensor, $\sigma^{\mu\nu}=i[\gamma^\mu,\gamma^\nu]/2$
is the antisymmetric Dirac tensor, and $\mu_{ij}$ is the neutrino magnetic moment matrix. In the general case, the $\mu_{ij}$ matrix can contain both diagonal and off-diagonal elements corresponding to transition magnetic moments. For Dirac neutrinos, both diagonal and transition magnetic moments are, in principle, allowed, whereas Majorana neutrinos cannot possess diagonal magnetic moments and can only have transition moments. Although transition magnetic moments are allowed in a general effective description, they introduce flavor-changing electromagnetic interactions. Since our aim is to probe new physics through the neutrino electromagnetic interaction with the minimal additional flavor structure, we consider only the diagonal magnetic moment elements. Among the diagonal flavor magnetic moments, those of the electron and muon neutrinos are already subject to stringent experimental constraints, whereas the diagonal magnetic moment of the tau neutrino remains considerably less constrained by direct measurements. Therefore, the numerical analysis presented in this work focuses exclusively on the diagonal magnetic moment of the tau neutrino, $\mu_{\tau\tau}$, while all other magnetic moment matrix elements are set to zero. 

\subsection{Signal Process and Equivalent Photon Approximation}
In this study, the sensitivity to the anomalous magnetic moment of the tau neutrino is investigated through the subprocess $\mu^{-}\gamma \rightarrow \mu^{-}\nu_{\tau}\bar{\nu}_{\tau}$, where the initial-state photon is described by the EPA \cite{vonWeizsacker:1934nji, Williams:1934ad}. Within this approach, a high-energy charged particle is surrounded by a flux of quasi-real photons with very small virtuality. Consequently, photon-induced interactions occurring in $\mu^{+}\mu^{-}$ collisions can be described with good accuracy as effective $\mu\gamma$ scattering processes. The EPA has become a standard framework for investigating photon-induced reactions at high-energy lepton and hadron colliders, as it delivers high accuracy while substantially reducing computational complexity.

\begin{figure*}[t!]
\centering
\includegraphics[width=13cm]{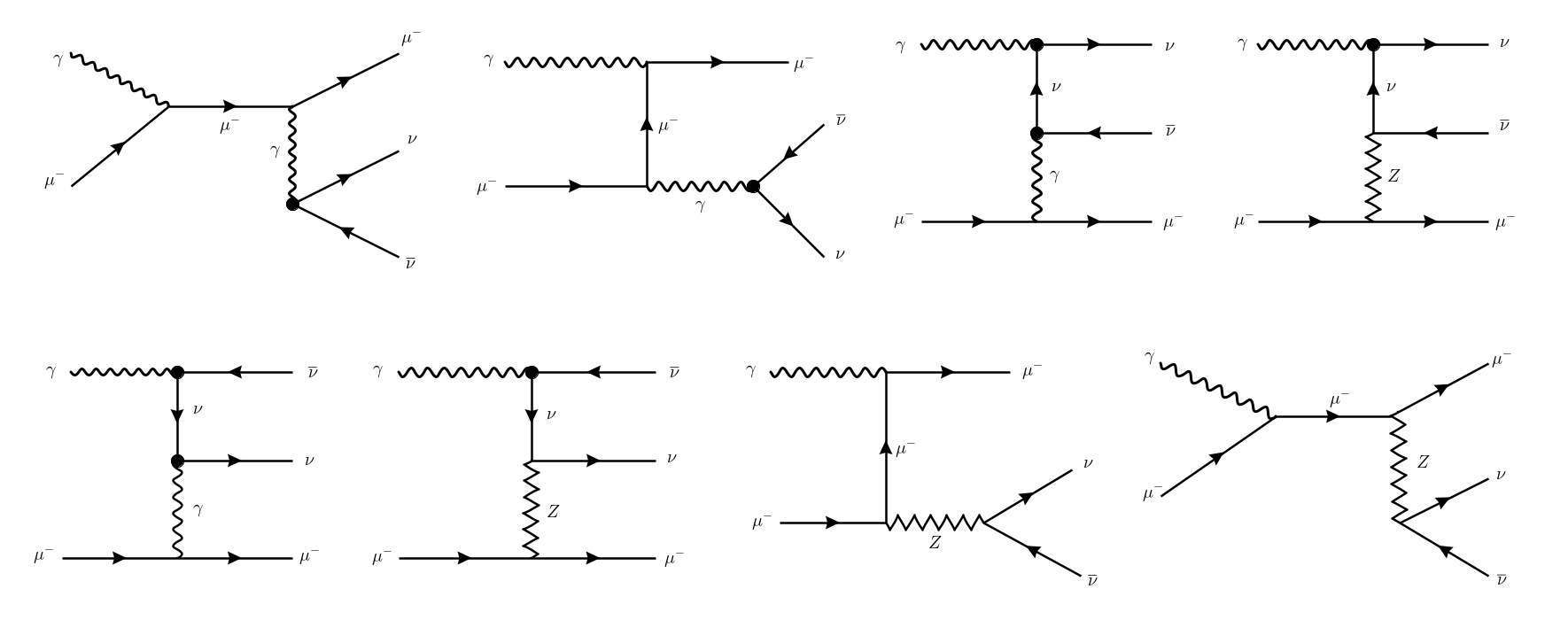}
\caption{Tree-level Feynman diagrams for the subprocess $\mu^{-}\gamma \rightarrow \mu^{-}\nu\bar{\nu}$ in the presence of nonstandard $\nu \overline{\nu} \gamma$ coupling marked with a black dot. }
\label{figure1}
\end{figure*}

Figure~\ref{figure1} presents a representative Feynman diagram for the subprocess
$\mu^{-}\gamma\rightarrow\mu^{-}\nu\bar{\nu}$. Owing to the clean collision environment, the relatively low level of hadronic backgrounds, and the high center-of-mass energies and luminosities foreseen for future muon colliders, this channel provides excellent sensitivity to the tau neutrino magnetic moment.

\begin{figure*}[t!]
\centering
\includegraphics[width=7cm]{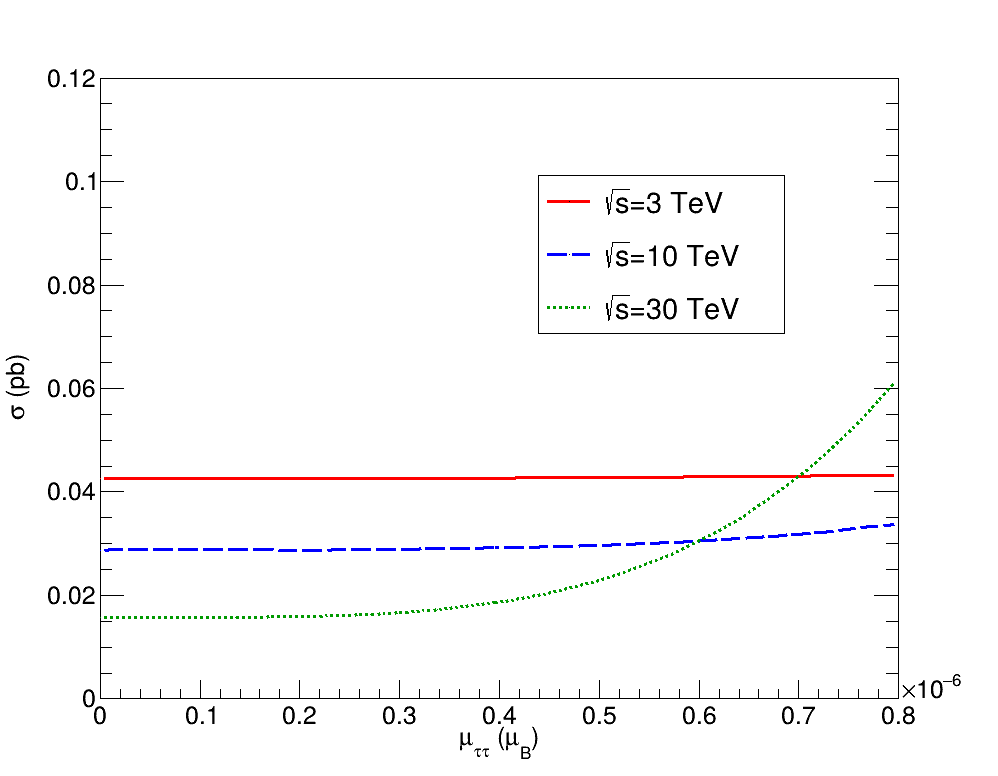}
\caption{The total cross-sections for the $\mu^- \gamma \rightarrow \mu^- \nu_{\tau} \bar{\nu}_{\tau}$ process as a function of the anomalous \(\mu _{\tau \tau}\) coupling for muon colliders with $\sqrt{s}=$ 3, 10 and 30 TeV. }
\label{figure2}
\end{figure*}

\section{Signal and Background Analysis}
\label{section3} 
High sensitivity to anomalous $\nu\bar{\nu}\gamma\gamma$ interactions through the $\mu^{-}\gamma \rightarrow \mu^{-}\nu\bar{\nu}$ process at future muon colliders has been demonstrated in a recent study \cite{DYilmaz:2026}. Motivated by this result, we investigate the potential of the same process to probe the anomalous magnetic moment of the tau neutrino. The signal and background processes considered in this analysis are classified as follows:
\begin{align*}
&\textbf{Signal:}\\
&\mu^{-}\gamma \rightarrow \mu^{-}\nu_{\tau}\bar{\nu}_{\tau}
\\[0.3cm]
&\textbf{Irreducible background:}\\
&\mu^{-}\gamma\rightarrow\mu^{-}\nu_l\bar{\nu}_l
\qquad (l=e,\mu,\tau)
\\[0.3cm]
&\textbf{Reducible backgrounds:}\\
&\mu^{-}\gamma\rightarrow\mu^{-}\gamma
\\
&\mu^{-}\gamma\rightarrow\mu^{-}q\bar q
\qquad (q=u,d,s,c,b)
\\
&\mu^{-}\gamma\rightarrow\mu^{-}l^-l^+
\qquad (l=e,\mu).
\end{align*}

The signal corresponds to the contribution of the anomalous magnetic moment interaction to the $\mu^{-}\gamma \rightarrow \mu^{-}\nu_{\tau}\bar{\nu}_{\tau}$ subprocess, whereas the irreducible background originates from the SM contribution to the same final state but for all neutrino flavors. The remaining channels constitute reducible backgrounds, since they may exhibit a comparable final state topology to that of the signal process, provided that their final states $ l^-$, $ l^+$, $\gamma$, and quarks are not detectable by the muon collider detectors.

The effective neutrino--photon interaction described by Eq.~(\ref{eq:Lmag}) was implemented in {\sc FeynRules} \cite{Alloul:2013bka} and exported in the Universal FeynRules Output (UFO) format \cite{Degrande:2011ua}. The resulting UFO model was then imported into {\sc MadGraph5\_aMC@NLO} \cite{Alwall:2014hca} for the calculation of the signal cross sections, and the generation of Monte Carlo event samples used throughout the analysis. Using this implementation, the dependence of the signal cross section on the anomalous magnetic moment parameter was conducted for the three muon collider center of mass energies. The corresponding dependence of the signal cross section on the anomalous magnetic moment parameter for the $\sqrt{s}=3$, 10, and 30 TeV muon collider options is presented in Fig.~\ref{figure2}. The events obtained at the Parton level were then subjected to Parton showering and hadronization processes using the PYTHIA8 program  \cite{Bierlich:2022pfr,Sjostrand:2014zea,Sjostrand:2019zhc}. To simulate realistic detector conditions, the PYTHIA8 outputs were run through a fast detector simulation using the DELPHES3 detector card developed for muon colliders \cite{MuonCollider:2022ded, Accettura:2023ked, Casarsa:2023vqx, Andreetto:2024rra, Black:2022cth, deFavereau:2013fsa}. At this stage, particle identification efficiencies, momentum and energy resolutions, detector acceptance, and reconstruction efficiencies were taken into account. Reconstruction of the final state particles and event analyses were performed using the MadAnalysis5 framework \cite{Conte:2012fm,Araz:2020lnp,Araz:2019otb,Dumont:2014tja,Conte:2014zja}.

Signal samples were generated for different benchmark values of the anomalous magnetic moment parameter. For each benchmark point, $5\times10^{5}$ events were generated in order to ensure sufficient Monte Carlo statistics for both the signal and all background processes. Consequently, the predicted cross sections and kinematic distributions are only minimally affected by statistical fluctuations.

In the pre-selection of signal and background events, trigger-level selection cuts were applied for the 3, 10 and 30 TeV operating points to ensure reliable triggering of final-state muons:
\[
P_T(\mu^-)>10~\mathrm{GeV},
\qquad
\slashed{E}_T >10~\mathrm{GeV}.
\]

In addition, the detector acceptance condition for the final-state muon was used in accordance with the muon collider detector design:
\[
|\eta(\mu^-)|<2.5
\]
\begin{figure*}[t!]
    \centering
    \includegraphics[width=0.43\textwidth]{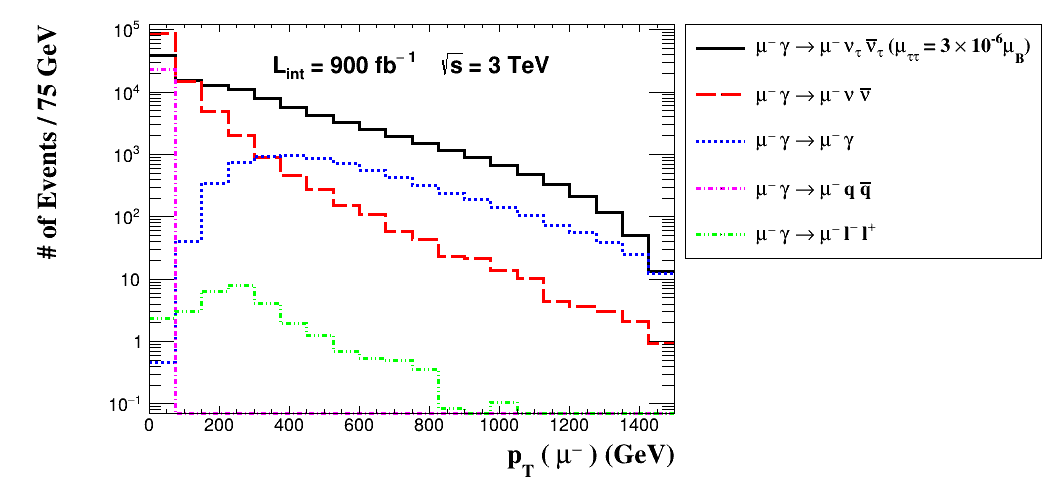} 
    \includegraphics[width=0.43\textwidth]{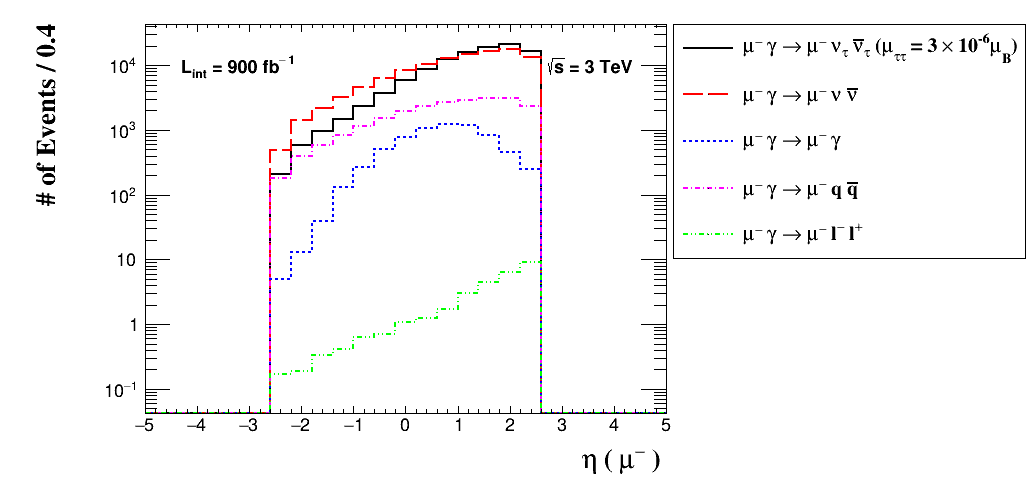}    
\vspace{0.35cm}
    \includegraphics[width=0.43\textwidth]{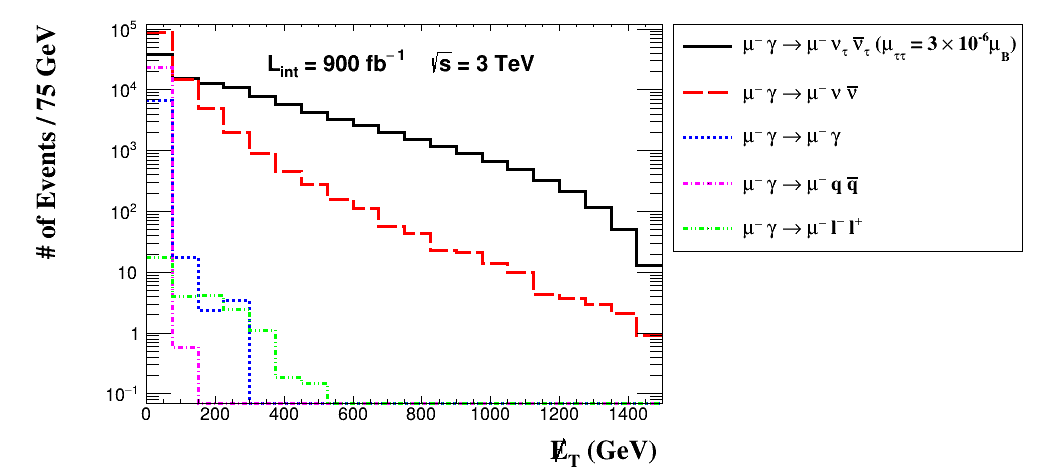}
    \caption{ The distributions of the transverse momentum $p_T(\mu^-)$, pseudorapidity $\eta(\mu^-)$, and missing transverse energy $\slashed{E}_T $ of the final-state muon are shown for the signal with $\mu_{\tau \tau} = 3.0 \times 10^{-6}\mu_B$ and the relevant background processes at $\sqrt{s}=3$ TeV.}
    \label{figure3}
\end{figure*}
For background processes containing hadronic final states, additional jet transverse momentum requirements were imposed in order to suppress contributions from soft QCD activity. Accordingly, jets were required to satisfy $35<P_T(j)<50~\mathrm{GeV}$ for the 3 and 10 TeV operating points, while the requirement was relaxed to $35<P_T(j)<150~\mathrm{GeV}$ for the 30 TeV collider option to account for the broader kinematic phase space at higher center of mass energy. After applying the pre-selection criteria, the transverse momentum ($P_T$), missing transverse energy ($\slashed{E}_T$), and pseudo-rapidity ($\eta$) distributions of the signal and background processes were examined in detail (Fig. \ref{figure3} - Fig.\ref{figure5}). Optimal selection cuts for each center-of-mass energy were determined based on these distributions. For $\sqrt{s}=3~\mathrm{TeV}$, we require $P_T(\mu^-)>600~\mathrm{GeV}$ and $\slashed{E}_T>150~\mathrm{GeV}$; for $\sqrt{s}=10~\mathrm{TeV}$, $P_T(\mu^-)>1500~\mathrm{GeV}$ and $\slashed{E}_T>500~\mathrm{GeV}$; and for $\sqrt{s}=30~\mathrm{TeV}$, $P_T(\mu^-)>4500~\mathrm{GeV}$ and $\slashed{E}_T>1500~\mathrm{GeV}$. These optimized cuts effectively suppress the backgrounds while largely preserving signal efficiency. In addition to these selection criteria, a veto condition was applied to eliminate events containing additional charged leptons in the final state: $N(e^-)=N(e^+)=N(\mu^+)=0$. The selection criteria were optimized to enhance the expected signal significance while retaining a sufficiently high signal acceptance. All selection criteria and corresponding cut values used in the final analysis are summarized in Table~\ref{table1}. 
\begin{figure*}[t!]
    \centering
    \includegraphics[width=0.43\textwidth]{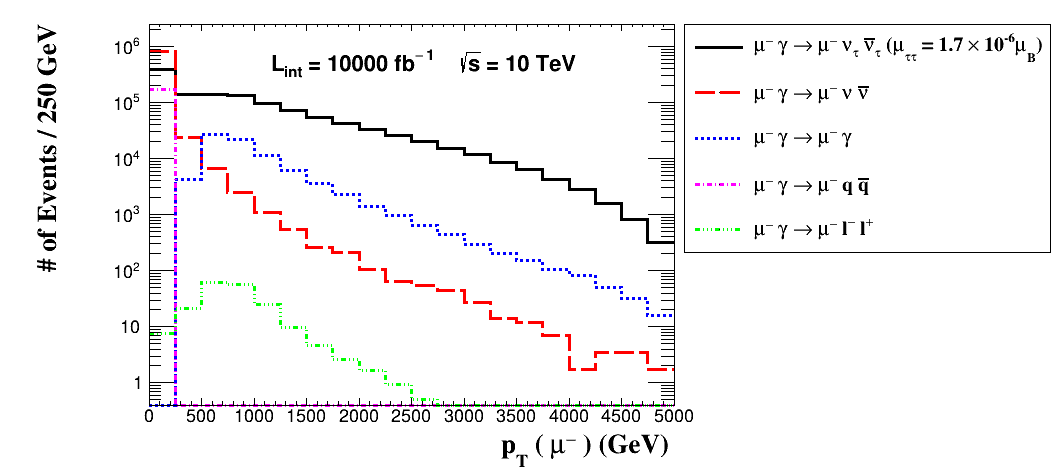} 
    \includegraphics[width=0.43\textwidth]{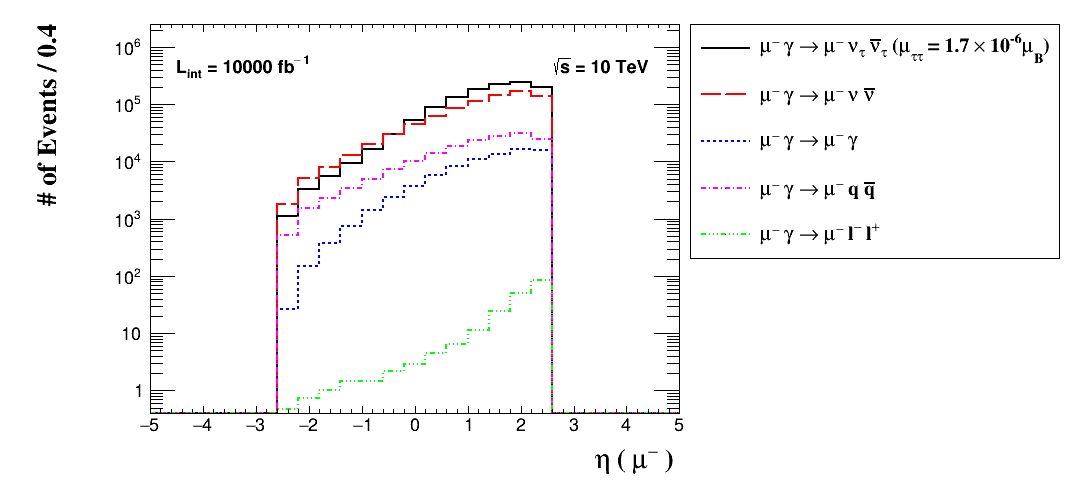}    
\vspace{0.35cm}
    \includegraphics[width=0.43\textwidth]{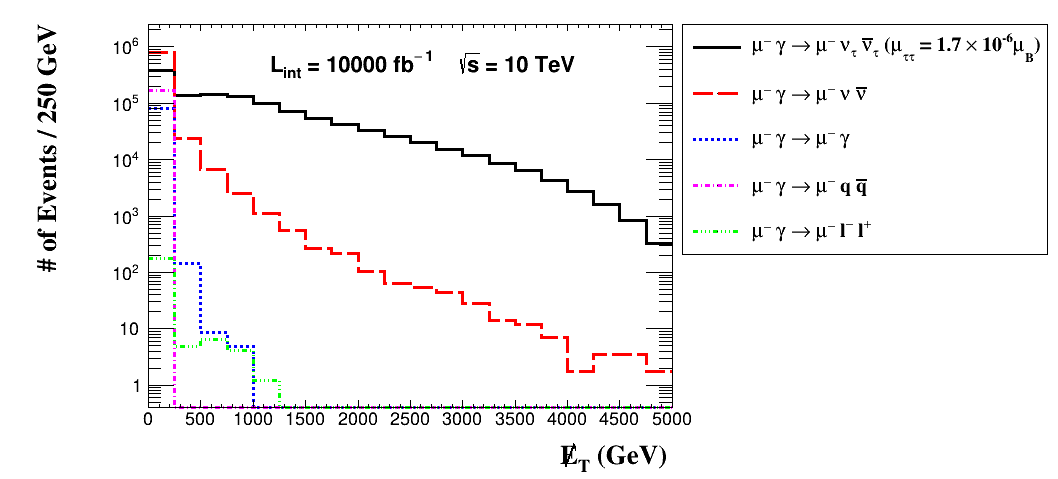}
    \caption{ The distributions of the transverse momentum $p_T(\mu^-)$, pseudorapidity $\eta(\mu^-)$, and missing transverse energy $\slashed{E}_T $ of the final-state muon are shown for the signal with $\mu_{\tau \tau} = 1.7 \times 10^{-6}\mu_B$ and the relevant background processes at $\sqrt{s}=10$ TeV.}
    \label{figure4}
\end{figure*}
\section{Sensitivity Calculations}
The sensitivity limits for the anomalous magnetic moment of the tau neutrino were determined using the Asimov significance method. Including the effect of systematic uncertainties, the Asimov significance is defined as follows \cite{Cowan:2010js, Elwood:2020pik}:
\begin{equation}
Z_A=
\left\{
2
\left[
(S+B)
\ln
\left(
\frac{(S+B)(B+\delta^{2})}
{B^{2}+(S+B)\delta^{2}}
\right)
-
\frac{B^{2}}{\delta^{2}}
\ln
\left(
1+
\frac{\delta^{2}S}
{B(B+\delta^{2})}
\right)
\right]
\right\}^{1/2}.
\end{equation}
\begin{table*}[b!]
\caption{Preselection and optimized event selection criteria used in the analysis. Here, $j$ denotes the quark-initiated jets corresponding to $u,\bar{u},d,\bar{d},s,\bar{s},c,\bar{c},b,\bar{b}$.}
\medskip
\centering
\begin{tabular}{llll}
\hline
\hline
\multicolumn{4}{l}{Preselection cuts}                                                                                  \\ \hline
$\sqrt{s}$= 3 TeV and 10 TeV & $P_{T}(\mu^{-}) > 10~\mathrm{GeV}$      &  $35 < P_{T}(j) < 50~\mathrm{GeV}$  & $\slashed{E}_T  > 10$ GeV \\
$\sqrt{s}$= 30 TeV           &  $P_{T}(\mu^{-}) > 10~\mathrm{GeV}$      &  $35 < P_{T}(j) < 150~\mathrm{GeV}$  & $\slashed{E}_T  > 10$ GeV \\
                          & $|\eta(\mu^-)|<2.5$ &                                  &                        \\ \hline
 \hline
\multicolumn{4}{l}{Selection cuts}                                                                                      \\ \hline
$\sqrt{s}$= 3 TeV            & $P_{T}(\mu^{-}) > 600~\mathrm{GeV}$      & $\slashed{E}_T  > 150$ GeV           &                        \\
$\sqrt{s}$= 10 TeV           & $P_{T}(\mu^{-}) > 1500~\mathrm{GeV}$      & $\slashed{E}_T  > 500$ GeV            &                        \\
$\sqrt{s}$= 30 TeV           & $P_{T}(\mu^{-}) > 4500~\mathrm{GeV}$      & $\slashed{E}_T  > 1500$ GeV            &                        \\
                          & $N(l)_{e^-, e^+, \mu^{+}}=0 $                    &                                  &                        \\ \hline
\end{tabular}
\label{table1}
\end{table*}
\begin{figure*}[t!]
    \centering
    \includegraphics[width=0.43\textwidth]{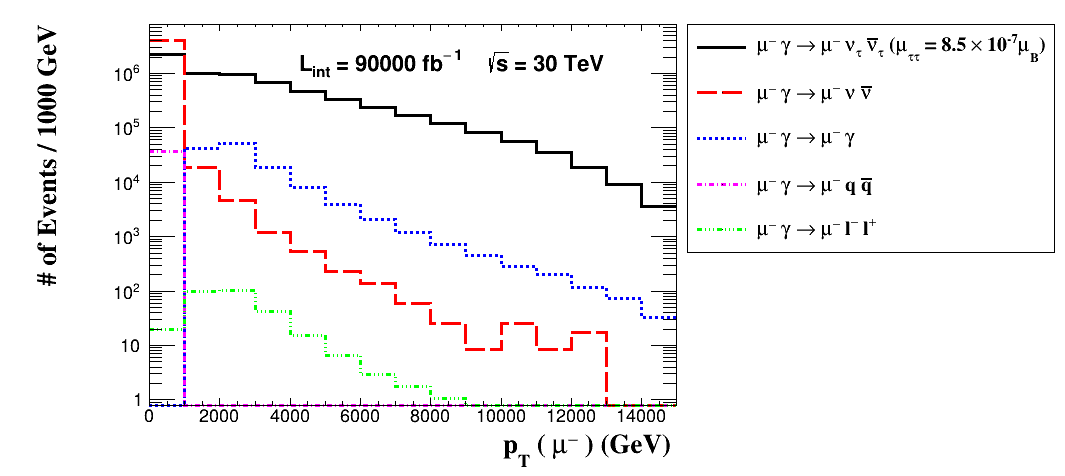} 
    \includegraphics[width=0.43\textwidth]{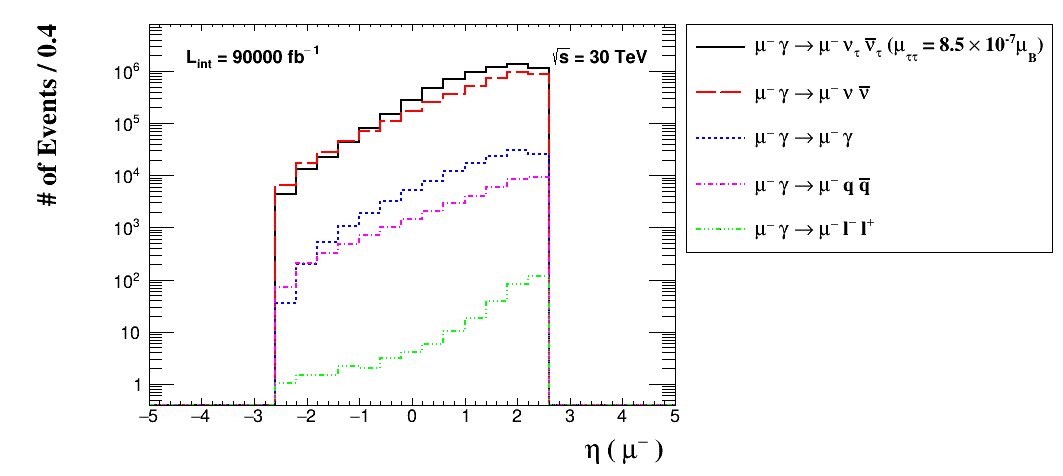}    
\vspace{0.35cm}
    \includegraphics[width=0.43\textwidth]{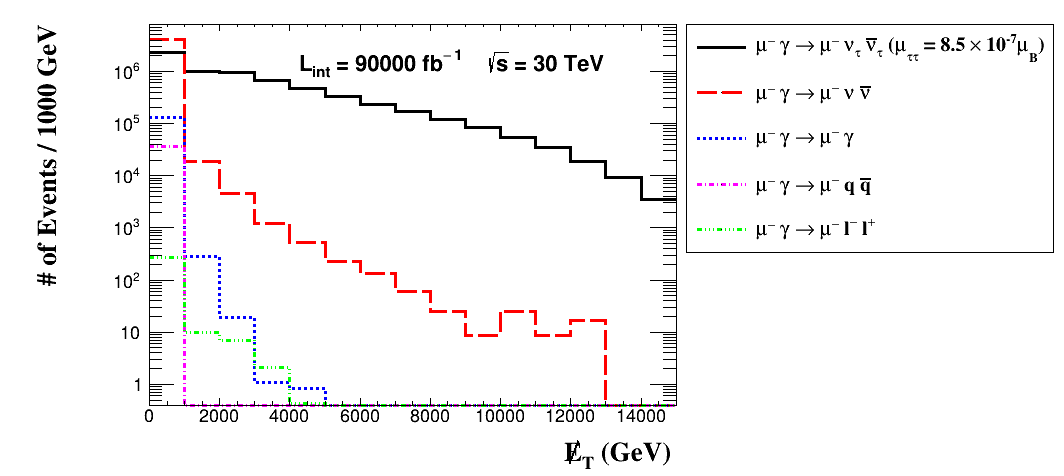}
    \caption{ The distributions of the transverse momentum $p_T(\mu^-)$, pseudorapidity $\eta(\mu^-)$, and missing transverse energy $\slashed{E}_T $ of the final-state muon are shown for the signal with $\mu_{\tau \tau} = 8.5 \times 10^{-7}\mu_B$ and the relevant background processes at $\sqrt{s}=30$ TeV.}
    \label{figure5}
\end{figure*}
Here, $S$ and $B$ represent the expected numbers of signal and background events, respectively, and $\delta$ denotes the assumed systematic uncertainty of the background. The sensitivity limits were determined by solving for the anomalous magnetic moment parameter at fixed significance levels of $5\sigma$, $3\sigma$, $1.64\sigma$, and $1.28\sigma$. The corresponding results for the 3, 10, and 30 TeV collider configurations are presented in Table~\ref{table2}, Table~\ref{table3}, and Table~\ref{table4}, respectively. 
The five-year integrated luminosities assumed for the three collider operating scenarios given in Table~\ref{table2} - ~\ref{table4} were calculated according to the following relation \cite{Accettura:2023ked, AlAli:2021let, Li:2023ksw}:
\begin{equation}
L_{\mathrm{int}}
=
10~\mathrm{ab}^{-1}
\left(
\frac{\sqrt{s}}
{10~\mathrm{TeV}}
\right)^2.
\end{equation}
\begin{table*}[t!]
\caption{Expected sensitivities (in units of $\mu_B$) to the anomalous magnetic moment of the tau neutrino at $\sqrt{s}=3$ TeV for $5\sigma$, $3\sigma$, $1.64\sigma$, and $1.28\sigma$ significance levels, obtained using the Asimov method.}
\begin{tabular}{|c|c|c|c|c|c|}
\hline
$\sqrt{s}$ (TeV) &$L_{int}$ ($\text{ab}^{-1}$)  &      & $\delta=1\%$ & $\delta=5\%$ & $\delta=10\%$ \\ \hline
\multirow{4}{*}{3} & \multirow{3}{*}{0.9} &5$\sigma$ & $7.52\times 10^{-7}$ & $8.88\times 10^{-7}$& $1.15\times 10^{-6}$         \\ \cline{3-6} 
                   &                      & 3$\sigma$ & $5.95\times 10^{-7}$& $6.70\times 10^{-7}$ &$8.60\times 10^{-7}$         \\  \cline{3-6} 
                   &                      & 1.64$\sigma$ & $4.88\times 10^{-7}$        & $5.21\times 10^{-7}$        & $6.65\times 10^{-7}$         \\  \cline{3-6} 
                   &                      & 1.28$\sigma$ & $4.60\times 10^{-7}$        & $4.82\times 10^{-7}$        & $6.11\times 10^{-7}$         \\ \hline
\end{tabular}
\label{table2}
\end{table*}

\begin{table*}[b!]
\caption{Expected sensitivities (in units of $\mu_B$) to the anomalous magnetic moment of the tau neutrino at $\sqrt{s}=10$ TeV for $5\sigma$, $3\sigma$, $1.64\sigma$, and $1.28\sigma$ significance levels, obtained using the Asimov method. }
\begin{tabular}{|c|c|c|c|c|c|}
\hline
$\sqrt{s}$ (TeV) &$L_{int}$ ($\text{ab}^{-1}$)  &      & $\delta=1\%$ & $\delta=5\%$ & $\delta=10\%$ \\ \hline
\multirow{4}{*}{10} & \multirow{3}{*}{10} &5$\sigma$ & $2.10\times 10^{-7}$ & $2.70\times 10^{-7}$& $3.73\times 10^{-7}$         \\ \cline{3-6} 
                   &                      & 3$\sigma$ & $1.79\times 10^{-7}$& $2.23\times 10^{-7}$ &$2.76\times 10^{-7}$            \\ \cline{3-6} 
                   &                      & 1.64$\sigma$ & $1.55\times 10^{-7}$        & $1.91\times 10^{-7}$        & $2.19\times 10^{-7}$         \\  \cline{3-6} 
                   &                      & 1.28$\sigma$ & $1.50\times 10^{-7}$        & $1.83\times 10^{-7}$        & $2.04\times 10^{-7}$         \\ \hline
\end{tabular}
\label{table3}
\end{table*}
The results demonstrate a clear improvement in sensitivity with increasing center-of-mass energy options of muon colliders. The 30 TeV collider configuration provides the strongest sensitivity, with the $5\sigma$ discovery reach of $\mu_{\tau\tau}=5.21\times10^{-8}\,\mu_B$ for a 1\% systematic uncertainty. At 10 TeV, the $5\sigma$ sensitivity reaches $2.10\times10^{-7}\,\mu_B$ for a 1\% systematic uncertainty, while at 3 TeV it is $7.52\times10^{-7}\,\mu_B$. These results demonstrate that future high-energy muon colliders can have great potential to probe the tau-neutrino magnetic moment.
\begin{table}[]
\caption{Expected sensitivities (in units of $\mu_B$) to the anomalous magnetic moment of the tau neutrino at $\sqrt{s}=30$ TeV for $5\sigma$, $3\sigma$, $1.64\sigma$, and $1.28\sigma$ significance levels, obtained using the Asimov method.}
\begin{tabular}{|c|c|c|c|c|c|}
\hline
$\sqrt{s}$ (TeV) &$L_{int}$ ($\text{ab}^{-1}$)  &      & $\delta=1\%$ & $\delta=5\%$ & $\delta=10\%$ \\ \hline
\multirow{4}{*}{30} & \multirow{3}{*}{90} &5$\sigma$ & $5.21\times 10^{-8}$ & $7.15\times 10^{-8}$& $1.06\times 10^{-7}$         \\ \cline{3-6} 
                   &                      & 3$\sigma$ & $4.12\times 10^{-8}$& $5.15\times 10^{-8}$ &$7.51\times 10^{-8}$         \\  \cline{3-6} 
                   &                      & 1.64$\sigma$ & $3.38\times 10^{-8}$        & $3.79\times 10^{-8}$        & $5.40\times 10^{-8}$  \\ \cline{3-6} 
                   &                      & 1.28$\sigma$ & $3.19\times 10^{-8}$        & $3.43\times 10^{-8}$        & $4.84\times 10^{-8}$  \\  \hline
\end{tabular}
\label{table4}
\end{table}
\section{Conclusion}

In this study, the sensitivity of future high-energy muon colliders to the anomalous magnetic moment of the tau neutrino was investigated within a model-independent effective field theory framework. Detailed Monte Carlo simulations including parton showering, detector effects, and optimized event selection criteria were performed for the proposed 3, 10, and 30 TeV muon collider operating scenarios. The expected sensitivities were evaluated using the Asimov significance method for different levels of systematic uncertainty and at several significance levels.

Under the single-parameter assumption adopted throughout this work, the diagonal matrix element $\mu_{\tau\tau}$ can be directly identified with the magnetic moment of the tau neutrino, allowing a comparison with existing experimental bounds. The strongest direct laboratory limit currently available was reported by the DONUT Collaboration \cite{DONUT:2001zvi},
\[
\mu_{\nu_\tau}<3.9\times10^{-7}\,\mu_B
\]
at the 90\% confidence level. At the 3 TeV operating point, the $1.28\sigma$ sensitivity reaches
\[
\mu_{\tau\tau}=4.82\times10^{-7}\,\mu_B
\]
for a 5\% systematic uncertainty, which is of the same order of magnitude as the current direct experimental bound. At higher energy options of muon collider, the sensitivity improves substantially; in particular, the 10 TeV and 30 TeV configurations probe the tau-neutrino magnetic moment up to one order below the current direct experimental limit.

On the other hand, most previous collider-based phenomenological studies have reported sensitivity limits on the tau-neutrino magnetic moment that are either weaker than or comparable to the current direct experimental bound established by the DONUT Collaboration. For instance, analyses based on photon--proton collisions at the LHC \cite{Sahin:2010zr, Sahin:2012zm} have achieved sensitivities of approximately $2\times10^{-6}\,\mu_B$, whereas photon--electron and photon--photon collision studies at CLIC \cite{Senol:2012sn}have reached limits of about $9\times10^{-7}\,\mu_B$. In contrast, the sensitivity limits obtained in the present analysis demonstrate that future high-energy muon colliders can substantially extend the sensitivity to the tau-neutrino magnetic moment.

In summary, this study demonstrates that future muon colliders offer excellent sensitivity to probe the anomalous magnetic moment of the tau neutrino. The combination of detailed Monte Carlo simulations, realistic detector modeling, and optimized event selection strategies indicates that the tau-neutrino magnetic moment can be probed down to the $\mathcal{O}(10^{-8})\,\mu_B$ level at a 30 TeV muon collider. These results provide strong motivation for including tau-neutrino electromagnetic interactions in the physics program of next-generation muon collider experiments.

\section*{Acknowledgment}
The numerical calculations reported in this paper were partially performed at TUBITAK ULAKBIM, High Performance and Grid Computing Center (TRUBA resources). This work was supported by the Scientific Research Projects Coordination Unit of Ankara University. Project Number: [FBA-2025-4234].


\end{document}